\documentclass[12pt]{article}

\usepackage[margin=1.0in]{geometry}
\usepackage{cite}

\usepackage{amssymb,latexsym}

\usepackage{epstopdf}

\usepackage{graphics,epsfig}

\usepackage{color}
\usepackage{xcolor}

\usepackage{amsmath,amsfonts}

\usepackage{mathrsfs}

\DeclareMathAlphabet{\mathpzc}{OT1}{pzc}{m}{it}

\usepackage{feynmp}
\usepackage{feynmp-auto}

\newcommand*{\Scale}[2][4]{\scalebox{#1}{$#2$}}%

\let\a=\alpha \let\b=\beta \let\g=\gamma \let\d=\delta \let\e=\epsilon
\let\z=\zeta \let\h=\eta \let\th=\theta  \let\k=\kappa
\let\l=\lambda \let\m=\mu \let\n=\nu \let\x=\xi \let\p=\pi %\let\r=\rho
\let\s=\sigma   \let\f=\phi  
 
      \let\G=\Gamma \let\D=\Delta \let\Th=\Theta \let\L=\Lambda
\let\X=\Xi  \let\S=\Sigma  \let\Y=\Psi
 
\let\la=\label  
  
\def\nn{\nonumber} \def\bd{\begin{document}} \def\ed{\end{document}}
\def\ds{\documentstyle} \let\fr=\frac \let\bl=\bigl \let\br=\bigr
\let\Br=\Bigr \let\Bl=\Bigl
\let\bm=\bibitem
\let\na=\nabla
\def\tU{{\widetilde U}}
\let\pa=\partial \let\ov=\overline
\def\ie{{\it i.e.\ }}
\newcommand{\be}{\begin{equation}}
\newcommand{\ee}{\end{equation}}
\def\ba{\begin{array}}
\def\ea{\end{array}}
\def\ft#1#2{{\textstyle{{\scriptstyle #1}\over {\scriptstyle #2}}}}
\def\fft#1#2{{#1 \over #2}}
\def\F#1#2{{ F_{#1}^{(#2)} }}
\def\cF#1#2{{ {\cal F}_{#1}^{(#2)} }}

\def\R{{\bf R}}
\def\sst#1{{\scriptscriptstyle #1}}
\def\oneone{\rlap 1\mkern4mu{\rm l}}
\def\e7{E_{7(+7)}}
\def\td{\tilde}
\def\wtd{\widetilde}
\def\im{{\rm i}}
\def\bog{Bogomol'nyi\ }
\newcommand{\ho}[1]{$\, ^{#1}$}
\newcommand{\hoch}[1]{$\, ^{#1}$}
\newcommand{\bea}{\begin{eqnarray}}
\newcommand{\eea}{\end{eqnarray}}
\newcommand{\ra}{\rightarrow}
\newcommand{\lra}{\longrightarrow}
\newcommand{\Lra}{\Leftrightarrow}
\newcommand{\ap}{\alpha^\prime}
\newcommand{\bp}{\tilde \beta^\prime}
\newcommand{\cB}{{\cal B}}
\newcommand{\cO}{{\cal O}}
\newcommand{\vecx}{\vec{x}}
\newcommand{\vecy}{\vec{y}}
\newcommand{\vecp}{\vec{p}}
\newcommand{\vecq}{\vec{q}}
\newcommand{\tr}{{\rm tr} }
\newcommand{\Tr}{{\rm Tr} }
\newcommand{\NP}{Nucl. Phys. }

\newcommand{\cL}{{\cal L}}
\newcommand{\cA}{{\cal A}}
\newcommand{\cT}{{\cal T}}
\newcommand{\cR}{{\cal R}}
\newcommand{\cD}{{\cal D}}
\newcommand{\cH}{{\cal H}}

\def\Cb{\bar{C}}

\def\sst#1{{\scriptscriptstyle #1}}
\def\0{{\sst{(0)}}}
\def\1{{\sst{(1)}}}
\def\2{{\sst{(2)}}}
\def\3{{\sst{(3)}}}
\def\4{{\sst{(4)}}}
\def\5{{\sst{(5)}}}
\def\6{{\sst{(6)}}}
\def\7{{\sst{(7)}}}
\def\8{{\sst{(8)}}}
\def\9{{\sst{(9)}}}
\def\p{{\sst{(p)}}}
\def\q{{\sst{(q)}}}
\def\ve{\varepsilon}
\def\vf{\varphi}
\def\F{\Phi}
\def\wg{\wedge}

\def\thb{\bar{\theta}}
\def\Thb{\bar{\Theta}}
\def\barp{\bar{p}}
\def\barq{\bar{q}}
\def\barc{\bar{c}}
\def\bard{\bar{d}}
\def\e{\epsilon}

\def \bi{\bibitem}
\def \la {\label}

\def \l {\lambda}
\def\foot{\footnote}
\def \tl  {{\tilde \l}}
\def \sql {{\sqrt \l}}
\def \adss {$AdS_5 \times S^5$\ }
\newcommand{\rf}[1]{(\ref{#1})}
\def \ov {\over}

\def\th{\theta}
\def\Th{\Theta}
\def\vth{\vartheta}
\def\btheta{{\bar\theta}}
\def\ttheta{{{\tilde\theta}}}
\def\bttheta{{{\bar\ttheta}}}
\def\vth{\vartheta}

\def\ra{\rightarrow}
\def\N{\nabla}
\def\F{{\cal F}}
\def\uM{\underline{M}}
\def\uA{\underline{A}}
\def\uN{\underline{N}}
\def\uP{\underline{P}}
\def\ua{\underline{a}}
\def\ub{\underline{b}}
\def\uc{\underline{c}}
\def\ud{\underline{d}}
\def\ue{\underline{e}}
\def\uf{\underline{f}}
\def\ui{\underline{i}}
\def\uj{\underline{j}}
\def\uk{\underline{k}}
\def\ul{\underline{l}}
\def\ual{\underline{\alpha}}
\def\ube{\underline{\beta}}
\def\um{\underline{m}}
\def\un{\underline{n}}
\def\up{\underline{p}}
\def\uq{\underline{q}}
\def\ur{\underline{r}}
\def\us{\underline{s}}
\def\umu{\underline{\mu}}
\def\unu{\underline{\nu}}
\def\ula{\underline{\l}}
\def\uka{\underline{\k}}
\def\usi{\underline{\s}}
\def\urh{\underline{\r}}
\def\cc{\circ}
\def\eqv{\equiv}

\def\ni{\noindent}

\def\Ep{E^{{}^{(+)}}}
\def\Em{E^{{}^{(-)}}}

\def\Mp{M^{{}^{(+)}}}
\def\Mm{M^{{}^{(-)}}}

\def \ha{{1\ov 2}}

\def\r{\rho}

\def\Y{{\rm Y}}
\def\X{{\rm X}}
\def\tY{\tilde{\rm Y}}
\def\tX{\tilde{\rm X}}
\def\dY{\dot{\rm Y}}
\def\dX{\dot{\rm X}}

\def \J {\mathcal{J}}
\def \del {\partial}

\def\dF{\dot{F}}
\def\dG{\dot{G}}
\def\df{\dot{f}}
\def \E {{\cal E}}
\def \S {{\cal S}}
\def \J {{\cal J}}

\def\ms{\mathcal{S}}
\def\mj{\mathcal{J}}
\def\soj{\fr{\ms}{\mj}}
\def \R {{\bf R}}
\def \om {\omega}
\def \bE {\bar E}
\def \x {{\cal X}}

\def \bi{\bibitem}
\def \la {\label}

\def \l {\lambda}
\def\foot{\footnote}
\def \tl  {{\tilde \l}}
\def \sql {{\sqrt \l}}
\def \adss {$AdS_5 \times S^5$\ }
\def \ov {\over}

\def \varpi {{\rm w}}

\def\thb{\bar{\theta}}
\def\Thb{\bar{\Theta}}
\def\mb{\bar{\m}}
\def\ab{\bar{\a}}
\def\zb{\bar{z}}
\def\psib{\bar{\psi}}
\def\barp{\bar{p}}
\def\barq{\bar{q}}
\def\barc{\bar{c}}
\def\bard{\bar{d}}
\def\e{\epsilon}
\def\wb{\bar{w}}
\def\lb{\bar{\l}}
\def\Jb{\bar{J}}
\def\Nb{\bar{N}}
\def\Zb{\bar{Z}}
\def\pab{\bar{\pa}}

\def\At{\tilde{A}}
\def\Bt{\tilde{B}}
\def\Ct{\tilde{C}}
\def\Dt{\tilde{D}}
\def\Et{\tilde{E}}
\def\Ft{\tilde{F}}
\def\Gt{\tilde{G}}
\def\Ht{\tilde{H}}
\def\Kt{\tilde{K}}
\def\Mt{\tilde{M}}
\def\Nt{\tilde{N}}
\def\Rt{\tilde{R}}
\def\at{\tilde{a}}
\def\bt{\tilde{b}}
\def\ct{\tilde{c}}
\def\dt{\tilde{d}}
\def\et{\tilde{e}}
\def\ft{\tilde{f}}
\def \ztt{\tilde{\z}}
\def \zetat{\tilde{\zeta}}
\def\htil{\tilde{h}}
\def\gt{\tilde{g}}
\def\nt{\tilde{n}}
\def\mut{\tilde{\mu}}
\def\nut{\tilde{\nu}}
\def\pht{\tilde{\f}}
\def\Phit{\tilde{\Phi}}
\def\vft{\tilde{\vf}}

\def\rht{\tilde{\rho}}

\def\asth{\hat{*}}
\def\phh{\hat{\phi}}

\def\bA{{\bf A}}

\def\ola{\overleftarrow}
\def\ora{\overrightarrow}
\def\alt{\tilde{\a}}

\def\eh{\hat{e}}
\def\eph{\hat{\e}}
\def\ph{\hat{p}}
\def\alh{\hat{\a}}
\def\beh{\hat{\b}}
\def\gah{\hat{\g}}
\def\Fh{\hat{F}}
\def\muh{\hat{\m}}
\def\nuh{\hat{\n}}
\def\thh{\hat{\th}}
\def\rhh{\hat{\r}}
\def\dh{\hat{d}}
\def\ih{\hat{i}}
\def\jh{\hat{j}}
\def\hh{\hat{h}}
\def\nh{\hat{n}}
\def\gh{\hat{g}}
\def\kh{\hat{k}}
\def\deh{\hat{\d}}
\def\wh{\hat{w}}
\def\lah{\hat{\l}}
\def\Ah{\hat{A}}
\def\Gh{\hat{G}}
\def\Kh{\hat{K}}
\def\Nh{\hat{N}}
\def\Rh{\hat{R}}
\def\Ch{\hat{C}}
\def\Omh{\hat{\Omega}}

\def\xh{\hat{x}}

\def\ps{\rlap{\, /}\;\,p }
\def\ks{\rlap{\, /}\;\,k }

\def\gym{g_{YM}}

\def\adot{\dot{a}}
\def\bdot{\dot{b}}
\def\bpa{\bar{\pa}}

\def\pr{\prime}
\def\ssk{\medskip}
\def\clb{\color{blue}}
\def\clr{\color{red}}
\def\clg{\color{green}}
\def\clp{\color{purple}}
\def\clc{\color{cyan}}
\def\clm{\color{magenta}}
\def\cly{\color{yellow}}

\def\bfA{{\bf A}}
\def\bfB{{\bf B}}
\def\bfK{{\bf K}}
\def\bfU{{\bf U}}
\def\bfX{{\bf X}}
\def\bfY{{\bf Y}}
\def\bfZ{{\bf Z}}
\def\bfg{{\bf g}}
\def\bfn{{\bf n}}

\def\bsk{\bigskip}
\def\ssk{\medskip}

\def\Ec{{\cal E}}

\begin{document}

\overfullrule=0pt
\parskip=2pt
\parindent=12pt
\headheight=0in \headsep=0in \topmargin=0in
\oddsidemargin=0in

\vspace{ -3cm}
\thispagestyle{empty}
%\vspace{1cm}
%\begin{flushright}
%Preprint DFPD 01/TH/\\
%hep-th/
%\end{flushright}

 \vspace{0.1cm}

\setcounter{equation}{0}
\setcounter{footnote}{0}
\setcounter{section}{0}

\begin{center}

{\Large\bf  Finite temperature contributions to cosmological 
	 constant
}

\vskip 0.8cm

\vspace{0.5cm}

I. Y. Park
\\

\vspace{0.3cm}

\vspace{0.3cm}
{\it Department of Applied Mathematics,
Philander Smith College %\footnote{Home institute}
                               \\
Little Rock, AR 72223, USA \\
inyongpark05@gmail.com
}

 \vspace{.5cm}

\end{center}

 \vspace{0.1cm}

\begin{abstract}

We reexamine the cosmological constant problem in a finite temperature setup and propose an intriguing possibility of carrying out perturbative analysis by employing a renormalization scheme in which the renormalized Higgs mass (or resummed mass, to be more precise) is taken to be on the order of the CMB temperature. Our proposal hinges on the fact that although the physical value of the cosmological constant does not depend on one's renormalization scheme, whether or not a fine tuning is involved does. The cosmological constant problem is avoided in the sense that the renormalization process no longer requires finetuning. This is achieved essentially by renormalization scheme-independence of a physical quantity, which in turn is assured by bare perturbation theory. The proposal shifts the cosmological constant problem to a peculiarity of the consequent perturbation series for the Higgs mass (and other massive sectors of the Standard Model); the peculiarity is interpreted as an indicator of new physics after the expected mathematical structure of the series is scrutinized. Finite-temperature-induced complexification of the effective potential is observed and its interpretation is given. A consistency check in the cosmology context is suggested.

\end{abstract}
\newpage

%%%%%%%%%%%%%%%%%

%\vspace{.3in}

%\ni {\bf Acknowledgments}

%\ni The research of this work was funded in part by Hangyang University, South Korea.

\section{Introduction}

The cosmological constant (CC) problem \cite{Weinberg:1988cp} (see, e.g., \cite{Padmanabhan:2002ji,Bianchi:2010uw,Martin:2012bt,Sola:2013gha} for reviews\footnote{The inspiring review \cite{Sola:2013gha} is largely based on the previous works by the author and his collaborators (see, e.g., \cite{Shapiro:1999zt,Shapiro:2000dz,Sola:2007sv}) that were further developed in \cite{Sola:2016jky,Peracaula:2018dkg,Moreno-Pulido:2020anb}. One of the main themes of these works is the so-called running vacuum. These works are further commented on in the conclusion and \cite{Park:2021vro}.}) arises from the loop effects of Standard Model (SM) particles, such as the Higgs particle: in the conventional renormalization schemes the contribution of those particles to the cosmological constant term is enormously larger than the observed value, thus leading to the CC fine tuning problem. Since the electroweak scale is much higher than, say, the temperature of the cosmic microwave background (CMB), it has been a standard practice to apply the zero-temperature setup to formulate and tackle the problem. However, the CC, as vacuum energy, is an infrared effect in the sense explained below, thus governed by the low energy sector. For this reason the infrared structure of the theory must be important and its meticulous description is desirable. In particular, the resolution of the problem would require renormalization of the vacuum energy with careful treatment of the infrared sector. The fact that the CC is a vacuum energy also implies that quantization and renormalization of {\em gravity} must be involved in its systematic treatment. In this work we show that the finite temperature effects with a new renormalization scheme in which the renormalization mass is taken to be small allow, when properly taken into account in the quantum gravitational setup, one to avoid the CC fine-tuning problem. The consistency of the new renormalization scheme is addressed. 

The proposed resolution does not, of course, directly predict the observed value of the CC; it allows one to avoid the {\em fine} tuning problem that should otherwise be involved in the renormalization program through which the physical value of the CC is determined - just as in any other coupling constant determination - with the experimental input. In other words, the proposed resolution makes the whole renormalization program more natural in that it is free of {\em fine}
tuning in the CC sector. Our proposal shifts the CC problem to "large-number tuning" of the Higgs mass sector and other massive sectors, if present. However, the latter is far more benign than the former, as we will explain in detail. Moreover the mathematical structure of the consequent
	series points toward novel and radical new physics regarding the hierarchy problem of the SM and the
	nature of physical mass of an elementary particle.

The vacuum energy is defined as a minimum of the effective potential. In vacuum energy computation both the ultraviolate (UV) and infrared (IR) structures play roles. To some extent the UV and IR contributions to the vacuum energy are intertwined. The relevance of the UV structure is evident since renormalization procedure is involved. The relevance of the infrared structure is subtler. It is well demonstrated in Casimir energy analysis (see, e.g., the account in \cite{Schwartz}). The point is that for proper evaluation of vacuum energy  it is necessary to employ an infrared regulator, such as a finite-size box in the Casimir case in momentum cutoff regularization. In the body we employ dimensional regularization with additional {\em finite} renormalization, and pay a close attention to the infrared structure introduced by the presence of the temperature. The finite temperature setup is not strictly necessary if one's sole objective is to avoid the CC problem: it is the new renormalization scheme in which the renormalized mass is taken to be small that plays an essential role in the avoidance. However, there is a question of how small is appropriately small, i.e., the question of the scales. The presence of the temperature provides a relevant scale, and makes the step of taking the small value less arbitrary than otherwise through optimal perturbation theory (OPT)\footnote{OPT is based on the variational principle. Other thermal physics techniques based on the variational principle include screened perturbation theory \cite{Karsch:1997gj}\cite{Andersen:2001ez} and 2PI formalism \cite{Blaizot:2000fc}.}  \cite{Stevenson:1981vj}. In other words, the setup provides a framework in which with experimental input the value of CC can be {\em naturally} determined in a renormalization procedure. Also, the setup must be generally relevant for various other cosmological contexts (see the discussion in the conclusion).

Are there indications that the finite temperature effects may a priori be important for CC analysis? Let us consider high and low temperature cases separately. Although high temperature is not directly relevant for the present analysis, there is something  utterly puzzling with its current status in the cosmological context. Finite temperature makes its contribution to the effective potential, in particular, to the CC (say, the $T^4$ term in \rf{1loopOPT} below). For a sufficiently high temperature, its contribution to the CC may well be the dominant component.\footnote{In particular, this raises another intriguing possibility that the early large CC may have an interesting implication for early dark energy porposals put forth to tackle the Hubble tension problem, as we further remark in the conclusion. \label{fnoede}} Furthermore, since the temperature should depend on time, so should the CC. This status of the matter notwithstanding, we are not aware of much literature where these issues at hand were given a serious consideration in the cosmological constant context. In our opinion the usual hydrodynamics analysis must be repeated with the finite temperature entries properly taken into account. One may raise that the effects of the temperature are taken into account in the usual hydrodynamics approach. However, one must make sure that all of the channels through which the temperature enters must be accounted for; the temperature contribution to the CC is not accounted for in the conventional hydrodynamics approach. Such a systematic study will have to address a subtle question of gauge symmetry breaking, and deserves separate work. This is one of the things that we are currently working on in \cite{Park}. The main study of the present work is the implication of the finite temperature effects in a low temperature regime; we leave the high-T discussion for now and come back in section 2, 3, and the conclusion. 

What about at a low temperature? Are there indications, despite the fact that the electroweak scale is much higher than the CMB temperature? First of all, it should definitely be possible, and is natural, to obtain zero-temperature results as a vanishing-temperature limit of the corresponding finite-temperature results. A hint of an indication comes from energy scalings in zero- and finite- temperature loop analyses. In zero temperature, a loop analysis typically yields logarithmic factors such as $\ln\fr{m}{\m}$, where $m$ is the renormalized mass of the field and $\m$ the renormalization scale. For the benefit of convergence, it is necessary to choose $\m\sim m$. (Or one may consider renormalization group equations to conduct certain resummations.) By the same token it will be desirable to take $\m\sim m \sim T$ once the temperature enters. The question is whether there is an additional, more quantitative rationale for enforcing the scaling. The answer is affirmative: in the present work this scaling is achieved in the course of improving the perturbative analysis by optimal perturbation theory as well as standard thermal resummation: we show that there is an OPT procedure that enforces the scaling.

In the body we reformulate the CC problem as a zero-temperature limit of the finite temperature counterpart. A potential obstruction to any finite-temperature perturbative analysis is the well-known infrared problem. (Reviews can be found in \cite{Kapusta,Bellac,Blaizot:2011va,Laine:2016hma,Senaha:2020mop}.) For a high temperature, say, that of the QCD era, a well-known example is the `Linde problem' \cite{Linde:1980ts}, which has been an active topic of research; see \cite{Ghiglieri:2020dpq}\cite{Du:2020odw} and references therein. An effective field theory approach combined with lattice computation was proposed to deal with the problem \cite{Braaten:1995cm}. The focus of the present work is a low temperature, the temperature of CMB. (Nevertheless, some of the textbook results obtained for a high temperature can be borrowed for reasons to be explained.) We show that the finite-temperature effects are in fact crucial for naturalness of taking a small renormalized mass: in particular, once the convergence property of the perturbation series is improved through a variant OPT, the optimized renormalized mass of the Higgs turns out to be essentially the temperature,\footnote{It has been revealed in the recent works of \cite{Park:2017dib,Nurmagambetov:2018het,Nurmagambetov:2020ann} that quantum corrections can qualitatively change the classical solution. In the present work we see a similar novelty: the finite temperature non-perturbative effects dictate, small temperatures notwithstanding, the renormalized mass.} thus addressing the CC fine-tuning problem at its root.

In essence, what is being proposed is an intriguing possibility of carrying out perturbative analysis by employing - with proper justifications - a renormalization scheme in which the renormalized Higgs mass is taken to be of the order of the CMB temperature. In a general renormalization scheme, there is hardly anything that stops one from adopting a renormalized mass of an arbitrary value - other than the motivation for such a choice. Bare perturbation theory is discussed and variant OPT is introduced to address this question. Also, our point of contention is that such a choice accommodates a wider range of physics and at the same time leads to interesting predictions in cosmology. More on this in section 2 and the conclusion. 

A cautionary remark is in order. The issue is that of renormalization schemes. There are two widely-used renormalization schemes: onshell and ``offshell" scheme, such as modified minimal subtraction ($\overline{\mbox{MS}}$). The former was predominantly used in the past whereas more recent literature employs the latter. (Among textbooks, the former is used in \cite{Weinberg}; many other textbooks, such as \cite{Sterman}\cite{Peskin}\cite{Schwartz}, employ the latter.) In the onshell scheme, the mass appearing in the Lagrangian is taken to be the physical mass whose more general definition is the pole of the 2-point Green’s function. In contrast, the mass in the $\overline{\mbox{MS}}$ scheme (or any scheme of renormalized perturbation theory,\footnote{The perturbative analysis employing an ``offshell" scheme often goes under the name of renormalized perturbation theory.} more generally) is a renormalized mass. The renormalized mass is initially viewed arbitrary: it gets determined {\em at the end} by requiring the pole of the Green’s function to match with the physical value, 125 GeV in the case of Higgs field. We take the renormalized mass (or the resummed mass, more precisely; see the body) of the Higgs to be of the order of the temperature, with the value of the physical mass intact as 125 GeV. We do so by making use of the freedom associated with choice of the value of the renormalized mass in renormalized perturbation theory. While the scaling argument above is a qualitative justification for the temperature-order renormalized mass, our OPT procedure provides a quantitative justification.

With the renormalized mass determined, the following task still remains: the zero-temperature theories, such as the zero-temperature Standard Model, have been quite successful. There, the renormalized masses turn out to be close to the pole masses, usually within a few percent. In the case of the SM Higgs field, for instance, the renormalized mass turns out close to 125 GeV, the pole mass value. If one now wants to take the renormalized mass to be around the CMB temperature, which is much smaller than the pole mass, one must yet maintain compatibility with the zero-temperature analysis: the resulting perturbation theory should preserve the success of the original zero-temperature theory. This is the main task that we undertake in section 2; briefly, what underpins the compatibility at a fundamental level is the fact that the theory is defined by the bare Lagrangian, which is what is behind renormalization scheme-independence of a physical quantity. In section 3, after examining vast freedom in choosing renormalization conditions, we invoke renormalization scheme-independence of physical quantities to affirm the compatibility. We note that a certain resummation is behind the invariance. (We also note that the small renormalized mass introduces a peculiarity in the renormalized perturbation theory series worth examining further for practical purposes.)

To be specific, the cosmological constant problem is examined by taking an Einstein-scalar with a Higgs-type potential. A variant optimal perturbation theory is implemented in the recently proposed quantum-gravitational framework. The optimized renormalized mass, i.e., the renormalized mass determined by the variant optimal perturbation theory, of the scalar field turns out to be on the order of the temperature. Since the CMB temperature is of the same order of magnitude as the measured value of the cosmological constant in today's Universe, the temperature-order CC implies that the cosmological constant problem is avoided. More rigorously, the temperature-order CC shifts the cosmological constant problem to compatibility of the consequent perturbative analysis. The compatibility is guaranteed essentially by invariance of physical quantities under renormalization scheme change, which is {\em assured by bare perturbation theory}. We point out the resummation behind the invariance.

In our OPT, finite-temperature-induced complexification of the effective potential is observed. In other words this effect is introduced through our variant OPT. It is thus additional to the well-known presence of the imaginary part in the zero-temperature effective potential. In particular, the vacuum expectation value of the scalar field itself becomes complex. We interpret the complexity to be instability - associated with the finite temperature - toward zero temperature.

With the renormalized mass being of the order of the temperature, the CC problem is greatly alleviated: the CMB temperature in terms of eV is $2.3\times 10^{-4}$ eV. Denoting this as $2.3 \times 10^{-13}$ GeV, the vacuum energy contribution associated with the thermal mass of a Higgs-type field is $\sim 10^{-51} \;\mbox{GeV}^{\,4}$. Given the fact that the rest of the fields in the SM will also contribute, which should compensate a few orders of magnitude, this appears within reasonable proximity of the observed CC value $\sim 10^{-47} \;\mbox{GeV}^{\,4}$.

\vspace{.2in}

The rest of the paper is organized as follows. Invariance of physical quantities under renormalization scheme change (as well as renormalization group flow) originates from the fact that the theory is defined by the bare Lagrangian. The invariance will play a crucial role in the proposed resolution of the CC problem. In section 2 we explore the freedom in adopting a renormalization scheme and do so for a zero-T setup to avoid inessential complications. We take a massive scalar theory with a quartic potential and discuss the perspectives of both bare and renormalized perturbation theories. The structure of the perturbation series is carefully examined and its potential implications are explored. In section 3 we carry out one-loop computation of the effective potential in a finite temperature setup. We do this for a Higgs-type scalar in a flat spacetime first, and subsequently extend the analysis to the graviton sector. By employing an OPT-induced renormalization scheme, we obtain the main result of the one-loop effective potential given in eq. \rf{mainres}. Since OPT generally improves perturbation theory, the fact that the new scheme is backed up by OPT makes it a choice that stands out among many generic schemes. The $T^4$-scaling obtained therein may raise a question since it is reminiscent of the scale parameter behavior of the radiation component in the $\L$CDM cosmology. This issue is addressed at the end of section 3.1 (and 4). In section 4 we end with summary, comments on potential implications of the present result for cosmology, and future directions.

%%%%%%%%%%%%%%%%%%%%%%%%%%%%%%
%%%%%%%%%%%%%%%%%%%%%%%%%%%%%%
\section{Freedom in renormalization scheme}
%%%%%%%%%%%%%%%%%%%%%%%%%%%%%%
%%%%%%%%%%%%%%%%%%%%%%%%%%%%%%

In this section we discuss two key conceptual aspects of our proposal. The first is its consistency. The renormalization scheme that we are employing in the main analysis is such that the Higgs renormalized mass is taken to be far smaller than the physical (i.e., pole) mass. Although the scheme is unconventional and unfamiliar, bare perturbation theory assures its legitimacy. For practical purposes it will nevertheless be useful to examine the new scheme's potential consistency issues. The second is the aforementioned peculiarity. Since the CC problem is so drastic a problem, one’s first impression could be that so “moderate" a measure as change of renormalization scheme could not possibly fit the bill. More thorough inspection, however, discloses that the measure is not so moderate as it may appear: there is a price to pay, a certain peculiarity. We will argue below that the peculiarity has the right balance: it is far-reaching but its peculiarity does not lead to a CC problem in another disguise. Instead, it points toward new physics.

The gist of the conventional analysis is as follows. After UV regularization one subtracts out the infinite part and fix the finite part of the vacuum energy. In $\overline{\mbox{MS}}$ scheme one removes essentially the $\fr{1}{\ve}$ part, and this amounts to fixing the finite part. At this point the renormalized mass is still undetermined; it is determined by matching the pole expression of the 2-point function with the physical value of the field, 125 GeV for a Higgs field. 

In section 3, we put forth an OPT that leads to the renormalized mass of the order of the temperature. With this one is to conduct the consequent perturbative analysis. In the proposed new scheme it is the value of the renormalized mass, instead of the finite part, that is fixed first, through our OPT, to be of the order of the temperature. Matching with the physical mass determines the finite part. This may raise a question on consistency of the resulting perturbative analysis, since the new renormalized mass is much smaller than the actual physical mass. Although the modification amounts, in nature, to finite renormalization and thus must not affect the physics, it will be useful to take a close look at how the perturbative analysis modifies in the new scheme.

The best formal framework for assuring the consistency is bare perturbation theory. (See, e.g., \cite{Hamada:2012bp} for a recent application of bare perturbation theory.) However, it will be useful to have a qualitative discussion of the reasoning behind the new scheme and its potential ramifications before quantitatively proving the consistency of the new scheme. Let us play with informal and intuitive modifications of the standard $\overline{\mbox{MS}}$ scheme of renormalized perturbation theory. We mainly consider a zero-T setup, since the issue at hand is already present at zero temperature and the crux of our proposal for avoiding the CC problem applies to the zero-T case as well. In the standard renormalized perturbation theory, say, with dimensional regularization and $\overline{\mbox{MS}}$ scheme, the finite part is pre-fixed and then the renormalized mass is determined at the end by matching the pole expression with the physical value of, say, 125 GeV. The renormalized mass turns out to be within a few percent of 125 GeV. For the sake of argument, let us suppose that the renormalized mass is determined to be 120 GeV. The precise value doesn't matter for the point to be made. Let us conduct the following `Gedankenexperiment' with the new scheme, i.e., the scheme in which the renormalized mass is fixed prior to the finite part denoted by $c_m$. (One of the purposes of the Gedankenexperiment is to show without the technicalities of bare perturbation theory that the new scheme points toward interesting predictions in the case of a finite-T setup; see below). One may take the renormalized mass $m$, say, $m=110$ GeV in the new scheme, and repeat the renormalization analysis. Since the difference between 120 GeV and 110 GeV is relatively minor, the perturbation series must remain valid and the finite constant $c_m$ will turn out to be close to the value prefixed in the standard approach. (Since $c_m$ is zero in MS scheme, its value in the new scheme will be small, namely, close to zero.) Now one keeps lowering the value of the renormalized mass $m$. One can be certain that the series will definitely be valid at least for a certain range of the renormalized mass values. The real question is whether or not one can modify the renormalized mass to a degree required to solve the CC problem. By employing bare perturbation theory we will see below that this is obviously the case. One can conduct a similar Gedankenexperiment involving a finite-T setup. Here one can play a similar game with two parameters (or three, including $c_m$), the renormalized mass and temperature. Let us consider the temperature to be around that of the electroweak (EW) era. The scheme leads to potentially very interesting predictions: the CC will turn out to be determined by the temperature, just the way it is determined in section 3. Then since the temperature should be time-dependent (checks are necessary but the time-dependence of the temperature in the expanding Universe should be generic), the result will lead to time-dependence of the CC.\footnote{Section 4.2 of \cite{Park:2016zgt} demonstrated the time-dependence of the CC in the zero-T setup.} More generally, one will have cosmology substantially different \cite{Park} from the standard one, as commented on in footnote \ref{fnoede}.

Let us turn to bare perturbation theory and assure that the key to preserving the success of the zero-temperature in the new scheme lies in vast freedom in choosing subtraction schemes and/or renormalization conditions. The system that we employ for this task is a real scalar theory with a quartic potential:
\bea
S= -\int d^4x\; \Big[\fr12 \pa_\m \z \pa^\m \z +\fr12 m^2 \z^2\Big]-\int d^4x\; \fr{\l}{4!}  \z^4. \la{esabe}
\eea
Invariance of the physical quantities under renormalization scheme change is based on the following well-known relation
\bea
{\cal L}_R(m,g\m^\e,c_m) = {\cal L}_{cl}(m_0,g_0)
\eea
where ${\cal L}_R$ (${\cal L}_{cl}$) denotes renormalized (classical) Lagrangian.	The theory is defined by the bare Lagrangian, i.e., the Lagrangian on the right-hand side. Let us again focus on the mass sector. Although one introduces $m$ and $c_m$, it is only a convenient ``splitting" of a {\em single} parameter $m_0$. One can obtain the pole mass expression to the desired order in bare perturbation theory. For example, let us consider
\bea
G_0(k,m_0)=\fr{1}{k^2+m_0^2-\Sigma(k,m_0)}
\eea
where $\Sigma(k,m_0)$ is a series in $\hbar$. The series is convergent; it is convergent irrespective of the aforementioned splitting of $m_0$. In general, a perturbation series in quantum field theory is an asymptotic one. However, the point to be stressed here is that whether the series is asymptotic or genuinely convergent does not depend on one's renormalization scheme, e.g., how one splits $m_0$: if the series under consideration is convergent for one renormalization scheme it will be so for other renormalization schemes as well.

To be entirely heuristic, let us consider the self-energy at one-loop; one finds 
\bea
\Sigma^\1=-\fr{\l}{32\pi^2}m_0^2 \Big(\fr{m_0^2}{4\pi \m^2}\Big)^{-\ve}\G(1-D/2).
\la{sigmaeq}
\eea
where $\ve\equiv \fr{4-D}2$ and $D$ denotes the spacetime dimension. To remove the divergence at this point, one can split $m_0$ into $m$ and the finite part $c_m$ according to one's choice of renormalization scheme. One can compute the series to a desired loop-order and subsequently split $m_0$. As this example clearly demonstrates, the splitting can be (quite arbitrarily) done {\em after} obtaining the series to a desired order without having to worry about the convergence. How the splitting is done is part of one's renormalization scheme, which amounts to redistributions of various values in the convergence series, i.e., certain resummation.  

Considering higher loops, the corresponding set of finite constants $c^{(n)}$ where $n$ denotes the loop order can be introduced as a matter of principle. Of course, it would be pleasant if things could be arranged to work with a `minimal' set of $c^{(n)}$, and this is what happens, e.g., in the SM analysis with the usual renormalization schemes. As we further comment below, one of the lessons of the present work is that one can cover a wider range of physics with a possibly slightly extended set of $c^{(n)}$ that are not prefixed.

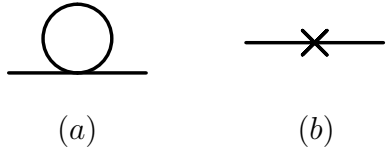
\begin{figure}
	\quad
	\centerline{
		\begin{minipage}[b]{10cm}
			\epsfxsize=12cm
			\[
			\begin{fmffile}{smcors2}
			\Scale[1.3]{
				\begin{gathered}
				\begin{fmfgraph*}(60,60)
				\fmfleft{i}
				\fmfright{o}
				\fmftop{m}
				%	\fmfv{label=$\tilde{\z}$,l.a=60}{i}
				%	\fmfv{label=$\tilde{\z}$,l.a=120}{o}
				%	\fmflabel{$\z$}{m}
				\fmf{plain,tension=1}{i,v1}
				\fmf{plain,tension=1}{v1,o}
				\fmf{plain,left,tension=0}{v1,m,v1}
				\end{fmfgraph*}
				\end{gathered}
			}
			\end{fmffile}
			\quad
			\begin{gathered}
			\begin{fmffile}{olctr}
			\Scale[1.3]{
				\begin{fmfgraph*}(40,60)
				\fmfleft{i}
				\fmfright{o}
				\fmftop{m}
				\fmfv{decor.shape=cross,decor.size=10}{v1}
				\fmf{plain,tension=1}{i,v1}
				\fmf{plain,tension=1}{v1,o}
				\fmf{phantom,left,tension=0}{v1,m,v1}
				\end{fmfgraph*}
			}
			\end{fmffile}
			\end{gathered} 
			\]
			\vspace{-.4in}	
			\[ (a)\hspace{1.0in}\;(b)\]
		\end{minipage}
	}
	\caption{Diagrams for one-loop self-energy: (a) one-loop correction for propagator (b) its counter-term}
	\label{fig1}
\end{figure}

%\[
%\begin{fmffile}{smcors2}
%\Scale[0.8]{
%	\begin{gathered}
%	\begin{fmfgraph*}(60,60)
%	\fmfleft{i}
%	\fmfright{o}
%	\fmftop{m}
%	%	\fmfv{label=$\tilde{\z}$,l.a=60}{i}
%	%	\fmfv{label=$\tilde{\z}$,l.a=120}{o}
%	%	\fmflabel{$\z$}{m}
%	\fmf{plain,tension=1}{i,v1}
%	\fmf{plain,tension=1}{v1,o}
%	\fmf{plain,left,tension=0}{v1,m,v1}
%	\end{fmfgraph*}
%	\end{gathered}
%}
%\end{fmffile}
%\quad
%\begin{gathered}
%\begin{fmffile}{olctr}
%\Scale[0.8]{
%	\begin{fmfgraph*}(40,60)
%	\fmfleft{i}
%	\fmfright{o}
%	\fmftop{m}
%	\fmfv{decor.shape=cross,decor.size=10}{v1}
%	\fmf{plain,tension=1}{i,v1}
%	\fmf{plain,tension=1}{v1,o}
%	\fmf{phantom,left,tension=0}{v1,m,v1}
%	\end{fmfgraph*}
%}
%\end{fmffile}
%\end{gathered} 
%\]
%\[
%\mbox{Fig. 1: diagrams for one-loop self-energy.  The cross denotes the counter-term for the one-loop two-point diagram.}
%\]

Although the bare perturbation theory component above assures the validity of our renormalization scheme, renormalized perturbation theory has its advantages in practical computations, thus it will be useful to examine the convergence issue in that framework. We do this by considering two-loop renormalization of the propagator. The two-point proper vertex is defined as
\be
\G^{[2]}\equiv k^2+m^2-\Sigma(k,m,c_m)
\ee
where $\Sigma$ again denotes self-energy. At one-loop, $\Sigma$ can be computed by considering the diagrams in Fig. 1: the one-loop two-point divergence introduces the counter-term: 
\bea
\begin{gathered}
	\begin{fmffile}{olctr}
		\Scale[0.8]{
			\begin{fmfgraph*}(40,60)
				\fmfleft{i}
				\fmfright{o}
				\fmftop{m}
				\fmfv{decor.shape=cross,decor.size=10}{v1}
				\fmf{plain,tension=1}{i,v1}
				\fmf{plain,tension=1}{v1,o}
				\fmf{phantom,left,tension=0}{v1,m,v1}
			\end{fmfgraph*}
		}
	\end{fmffile}
\end{gathered}\;
:\;-\fr{m^2}{4}\fr{\l}{(4\pi)^2}\;\Big(\fr{1}{\ve}+c_m\Big)\z^2
\la{mscmctr}
\eea
where $\ve\equiv \fr{4-D}2$ as before and $D$ denotes the spacetime dimension; $c_m$ is to be determined by one's subtraction scheme. For instance, the modified minimal subtraction ($\overline{\mbox{MS}}$) corresponds to setting $c_m=0$. Fixing the renormalized mass according to the OPT principle of minimal sensitivity\footnote{The idea behind principle of minimal sensitivity is that the best approximation is the one which is least sensitive to a small change in the renormalization scheme, since the full result must be independent of the scheme \cite{Stevenson:1981vj}.} (PMS) (as we will in section 3) is in contrast to the usual practice in zero-temperature: there, one fixes $c_m$ by one's subtraction scheme, and then the renormalized mass by the pole mass condition, 
\be
k^2+m^2-\Sigma(k)\,\Big|_{k^2=-m_P^2}=0
\ee 
where $m_P$ denotes the physical pole mass. In the new scheme, it is the coefficient $c_m$ that is determined by the pole mass condition, while the renormalized mass is prefixed (by variant OPT in section 3).

As stated in the introduction, the advantage of the new scheme is obvious: it realizes the scaling mentioned in the introduction and thereby allows one to avoid the CC fine-tuning problem. To see the disadvantage, let us note that the pole mass condition in the new scheme yields 
\be
c_m\, m^2 \sim m_P^2-m^2  \la{tde}
\ee
and thus implies a larger value of $c_m$, compared with the standard approach where the pole mass condition typically leads to $m^2\simeq m_P^2$. Note, however, that it is the combination $m^2 c_m \sim m_p^2$, but not $c_m$ alone, that appears in the counter-term \rf{mscmctr}. In general, for mass-related quantities it will not be possible at tree-level to achieve suitable agreement with experimental values, since the renormalized mass is pre-fixed: it will be necessary to go to one-loop where one has the freedom of adjusting the finite parts. The two-loop-relevant diagrams are given in Fig. 2. The circle in the last diagram in Fig. 2 represents the counter-term for the one-loop four-point diagram (not explicitly shown). Adding all up, the total two-loop self-energy $\Sigma^\2$ is
\bea
\Sigma^\2
=m^2\fr{\l^2\m^{4\ve}}{(4\pi)^4}
\Big[\fr{1}{2\ve^2}+\fr{1}{4\ve}(-1+c_m +3c_\l)
-\fr{p^2}{24m^2\ve}+\cdots\Big]
\eea
where $c_\l$\footnote{The new scheme concerns only the renormalized mass: for the other quantities, such as the coupling constants, one can just employ the standard schemes.} denotes the finite part of the counter-term of the one-loop four-point amplitude and $p$ the momentum entering through one end of the diagrams. The $\ve$-pole terms will have to be removed by two-loop counter-terms. Let us focus on the finite parts. The constant $c_m$ - $m^2 c_m$, more precisely - will also appear in the finite parts represented by the ellipses.\footnote{Since the issue under consideration belongs to the mass, one may well set $c_\l=0$ in the spirit of the $\overline{\mbox{MS}}$. Similarly, the two-loop analogues of $c_m, c_\l$ (and the finite part associated with the wavefunction renormalization) may be set to zero. In other words, too much freedom in choosing the finite parts can be a burden: the freedom remaining after determining $c_m$ can be fixed just as in a convenient subtraction scheme, such as the $\overline{\mbox{MS}}$, to facilitate `next' steps, e.g., solving the renormalization group equations.} By imposing the pole mass condition and solving it, say, interactively for $c_m$, it will be possible to determine its two-loop correction. Since $\Sigma^\2$ has an additional $\l$ and $\hbar$ compared with one-loop, the presence of $m^2 c_m$ cannot disturb the series in any significant way. (It is bare perturbation theory that ultimately confirms this.) After all, the new perturbation can be viewed as finite renormalization. We will point out later that a certain resummation is behind this finite renormalization.

%Let us push the logic in the Gedankenexperiment above to a more formal level. In the standard approach, one matches various physical values with the corresponding theoretical expressions to determine the renormalized values in terms of the physical parameters. While doing so, one treats the renormalized quantities, and also the finite parts in a more general procedure, as the unknowns. Essentially the same steps apply to the new scheme. Focusing on the mass sector, one may take the unknowns as $m,\Sigma$ instead of $m,c_m$ when matching with the physical quantities. With $m,\Sigma$ (or $m,c_m$, equivalently) being arbitrary initially, one can perform the calculation implicitly with  $\Sigma, m$, without being hampered by the convergence issue, which is also what one does in the standard analysis.

\begin{figure}
	\quad
	\centerline{
		\begin{minipage}[b]{10cm}
			\epsfxsize=12cm
			\[
			\begin{fmffile}{twopointloop}
			\Scale[.9]{
				\begin{gathered}
				\begin{fmfgraph}(40,60)
				\fmfleft{i1,i2,i3}
				\fmfright{o1,o2,o3}
				\fmf{plain}{i1,v1,o1}
				\fmffreeze
				\fmf{phantom}{i2,v2,o2}             
				\fmf{phantom}{i3,v3,o3}
				\fmf{plain,left,tension=.3}{v1,v2,v1}
				\fmf{plain,left,tension=.2}{v2,v3,v2}
				%	\fmfdot{v1,v2}
				\end{fmfgraph}
				\end{gathered}
			}
			\end{fmffile}
			\quad\quad\begin{fmffile}{sunset}
			\Scale[1.2]{
				\begin{gathered}
				\begin{fmfgraph}(40,40)
				\fmfleft{i}
				\fmfright{o}
				\fmf{plain,tension=5}{i,v1}
				\fmf{plain,tension=5}{v2,o}
				\fmf{plain,left,tension=0.4}{v1,v2,v1}
				\fmf{plain}{v1,v2}
				%	\fmfdot{v1,v2}
				\end{fmfgraph}
				\end{gathered}
			}
			\end{fmffile}
			\quad\quad\begin{fmffile}{tlwc3}
			\Scale[1.2]{
				\begin{gathered}
				\begin{fmfgraph*}(40,60)
				\fmfleft{i}
				\fmfright{o}
				\fmftop{m}
				\fmfv{decor.shape=cross,decor.size=10}{m}
				\fmf{plain,tension=1}{i,v1}
				\fmf{plain,tension=1}{v1,o}
				\fmf{plain,left,tension=0}{v1,m,v1}
				\end{fmfgraph*}
				\end{gathered}
			}
			\end{fmffile}
			\quad\quad\begin{fmffile}{tlwc4}
			\Scale[1.2]{
				\begin{gathered}
				\begin{fmfgraph*}(40,60)
				\fmfleft{i}
				\fmfright{o}
				\fmftop{m}
				\fmfv{decor.shape=circle,decor.size=5}{v1}
				\fmf{plain,tension=1}{i,v1}
				\fmf{plain,tension=1}{v1,o}
				\fmf{plain,left,tension=0}{v1,m,v1}
				\end{fmfgraph*}
				\end{gathered}
			}
			\end{fmffile}
			\]
			%	\vspace{-.2in}	
			%\[ (a)\hspace{1in}(b)\]
		\end{minipage}
	}
	\caption{Diagrams for two-loop self-energy}
	\label{fig2}
\end{figure}
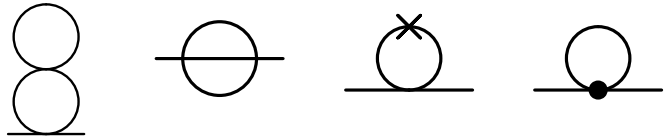

To be thorough, let us take the counter-term, $\sim c_m^\1 m^2 \z^2$ and examine its feedback to the CC in renormalized perturbation theory to see how things are organized in this sector. (Note that we have now denoted $c_m$ by $c_m^\1$ to distinguish the two-loop finite parts from those at one-loop.) The counter-term $c_m^\1 m^2 \z^2$ will yield, upon self-contracting the $\z$'s, a contribution to the CC at two-loop. Such a contribution does not cause any harm to the two-loop analysis in \cite{Park:2021vro}, as we now show. The contraction, being divergent, necessitates the corresponding CC renormalization at two-loop by introducing a CC counter-term $\delta \Lambda$. The renormalization procedure introduces $c_\Lambda^\2$. It is natural to take  $c_\Lambda^\2$ to be small. Needless to say that an additional factor of $\hbar$ dampens the two-loop result. One must choose its value in a manner consistent with the one-loop result and known CC value. The presence of the counter-term implies that it will be useful to carry out the resummation analogous to the one invoked when treating the mass term of a massive theory as a vertex. As well known, one gets the usual massive propagator after resumming the mass vertex corrections to the massless propagator. The resummation needed for the present case will of course be the finite-T analogue.

The discussion above illustrates that the way the CC problem manifests itself clearly depends on one's renormalization scheme. Needless to say, it also depends in general on the regularization method. As stated in the abstract, the proposed resolution shifts the cosmological constant problem to a peculiarity in the consequent perturbation series for the Higgs mass (and other massive sectors of the Standard Model). We now argue that it can be interpreted as an indicator of new physics.\footnote{ Before getting to the interpretation, let us first examine the expected mathematical structure of the series more carefully. As noted around \rf{tde} it {\em appears} that the one-loop correction has to be large to match the physical mass of the Higgs. It also seems that one would encounter even greater peculiarity in other massive sectors, if present, since one is unlikely have a luck of getting analogues of the $m^2 c_m$ combination. These expectations are somewhat naive because it has been {\em assumed} that the dominant piece should be one-loop with the higher loops being far smaller. Although this is usually the case in conventional schemes, this pattern may not necessarily be true in the present scheme. It is the total sum of the quantum corrections that must be matched to the physical mass. In other words, it may well be that the series is {\em slowly} convergent. To convey the point, suppose, for the sake of the argument, that it is possible to obtain the sum in a closed form which must be a highly complex function of $c_m$. (In practice, one will have to calculate the two point functions up to sufficiently high order. To precisely which order will be determined by the behavior of $c_m$ as a function of  loop order $n$, $c_m(n)$. It is the order of $n$ on which $c_m(n)$ does not substantially change that determines the order $n$; one must go to that order for proper evaluation of $c_m$.) The finite constant $c_m$ will be determined by matching the closed form with the physical mass, and it is only the closed form result that can reveal the true size of $c_m$. Only then one can retrospectively check relative sizes of the terms in the series. There might well be a ``conspiracy" of many terms, as opposed to a few dominant terms, that brings the value of the physical mass. Although the Higgs mass makes the largest contribution to the CC in the onshell scheme, there are other massive sectors wherein analogues expressions of $m^2c_m$ may not arise. Again, our contention is that the series may well be slowly convergent ones. Regardless, the peculiarity or ``large-number tuning" seems far benign than the fine-tuning problem. (More in the conclusion.) \la{fn9}}

In the original problem fine tuning was required since the "natural value" of the CC far exceeded the observed value: one must reach the physical value by finely adjusting the tiny decimal part of the subtracted number. In the present case, however, the physical mass is large, unlike the physical value of the CC. The necessary adjustment of $c_m$ should be on its leading part, thus is not {\em fine} tuning. One may still view the adjustment procedure a tuning of some sort. To our view, however, it is not too different from a typical parameter-determination procedure in renormalization program. In the conclusion we suggest some future useful checks to better settle this.

Let us get to the interpretation of the peculiarity. The present proposal has an interesting, though speculative, implication for the hierarchy problem. It lessens its severity: in spite of the very different mass scales in the SM, there exists a common link: the very temperature scale. One can repeat the analysis in this sector in the other massive sectors of the SM: it is not the physical values of the particles but the finite constant $c$ for each of them that is a primary parameter. We take up several other ramifications of our proposal in the conclusion.

%%%%%%%%%%%%%%%%%%%%%%%%%%%%%%
%%%%%%%%%%%%%%%%%%%%%%%%%%%%%%
\section{Variant OPT-induced renormalized mass}
%%%%%%%%%%%%%%%%%%%%%%%%%%%%%%
%%%%%%%%%%%%%%%%%%%%%%%%%%%%%%

This section aims to establish the naturalness of taking a small renormalized mass in the finite temperature setup. The crux of the variant optimal perturbation theory (OPT) can be captured by considering a scalar system in a flat spacetime. Since the UV divergences originate locally from a short distance, they are insensitive to global geometry. For this reason, the zero-temperature UV regularization can be employed in a finite-temperature theory. As for quantities depending on the infrared structure, the prime example of which is vacuum energy, one must consider in principle the actual background. The difference between using the curved background and the flat one lies in the finite parts. (However, the finite parts are adjusted by the renormalization conditions anyway; we refer to \cite{Park:2018vci} for further discussion.)

In thermal field theory, convergence of perturbative analysis is improved by resummation. The convergence can be further enhanced with a touch of non-perturbative techniques, such as OPT. The OPT implemented in this work is a relatively minor, but nonetheless crucial, variation of the widely-studied one. In the widely-used OPT, an additional artificial mass term is subtracted out after adding. This is one way of ensuring artificial-mass independence of the full closed results. In our case, the renormalized mass itself serves as the OPT parameter to be fixed by the OPT principle of minimal sensitivity. As well known in the context of the variational principle in quantum mechanics, there is no unique way of implementing the principle. For instance, the more variational parameters one introduces, the more accurate the approximation generally becomes. Our OPT is one that has an advantage of achieving the goal of avoiding the CC problem through improved infrared convergence.

%\subsection{Finite-temperature renormalization of CC }

\vspace{.2in}

The gravity-scalar system that we consider is
%%%
\bea
S=\fr1{\k^2}\int d^4 x \sqrt{-g}\;R  - \int d^4 x\sqrt{-g}\; \Big(\fr12g^{\m\n}\pa_\m\z \pa_\n \z +V(\z)\Big)
\la{grv-sclr}
\eea
%%%
where $\k^2= 16\pi G$ with $G$ being Newton's constant.
The potential $V(\z)$ is 
\bea
V(\z)=\fr{\l}{4!}\Big(\z^2+\fr{6}{\l} \n^2\Big)^2.  \la{mzpot}
\eea
Note the notation change as compared with section 2: the mass has been denoted by $\n$. One conceptual hurdle is the justification of the complete-square form of the potential instead of the usual $V=\fr12 \n^2 \z^2+\fr1{4!} \l \z^4$. The value of CC depends, of course, on whether one uses the complete-square form or the form without the constant piece. A shift of potential by a constant is immaterial in flat spacetime quantum field theory. The same is not true, however, in the quantum gravitational context. Whether one should use the complete-square form or the more usual form is not part of the CC problem. A closely related question, whose answer is not currently known \cite{Weinberg}, is why the minimum value of the classical Higgs potential should be taken to be zero. It is an independent problem that must ultimately be answered experimentally. Our goal here is to show that in the setup dictated largely by convergence of thermal perturbation theory, the fine tuning-problem is not present; this goal can be achieved more conveniently with the complete-square form.

To set the stage for the refined BFM \cite{Park:2014noa,Park:2016zgt,Park:2018vci,Park:2019amz}, we shift the fields as
\bea
g_{\m\n}\ra h_{\m\n}+\tilde{g}_{\m\n}\quad,\quad
\z \ra \hat{\z}+\tilde{\z}
\la{gshift}
\eea
with
\be
\tilde{g}_{\m\n}\equiv g_{c\, \m\n}+\vf_{\m\n}\quad,\quad \tilde{\z}\equiv \z_c+\xi
\ee
where $g_{c\, \m\n},\z_c$ denote the classical solutions, $\vf_{\m\n},\xi$ the background fields, and $h_{\m\n}, \hat{\z}$ the fluctuation fields.
The zero-temperature loop analysis is based on the following two-point functions (see \cite{Park:2019amz} for the conventions). For the metric,
\bea 
<h_{\m\n}(x_1)h_{\r\s}(x_2)>&=& \tilde{P}_{\m\n\r\s}\, \tilde{\D}(x_1-x_2)  \la{h2pt}
\eea
%%%
where the tensor $\tilde{P}_{\m\n\r\s}$ is given, in de Donder gauge, by \cite{Park:2014noa}
%%%
\bea
\tilde{P}_{\m\n\r\s} &\equiv& \fr{\bar{\k}^2}2\Big(\gt_{\m\r}\gt_{\n\s}+\gt_{\m\s}\gt_{\n\r}
- \fr12\gt_{\m\n}\gt_{\r\s}\Big);   \la{fpt}
\eea
%%%
where $\bar{\k}^2\equiv 2\k^2$ and satisfies
\bea
\tilde{P}_{\m\n\k_1\k_2}\tilde{P}^{\k_1\k_2}{}_{\r\s}=\bar{\k}^2\tilde{P}_{\m\n\r\s}. \la{Pprop}
\eea
$\tilde{\D}(x_1-x_2)$ is Green's function for a (massless) scalar theory in the background metric $\gt_{\m\n}$:
\bea 
<\hat{\z}(x_1)\hat{\z}(x_2)>&=&  \tilde{\D}(x_1-x_2).   \la{s2ptg}
\eea 
The explicit form of $\tilde{\D}$ for a massive scalar theory will be given and utilized in section 3.2 where the curved space analysis of the matter-involving sector is conducted. With these the finite temperature can be computed in the standard way.

In passing, the following can be said for breaking of gauge invariance in a finite temperature setup. In gravity theory, introducing a finite temperature should be viewed as part of specifying the background. Once temperature enters, the diffeomorphism is lost in a manner that can be dubbed as generalized spontaneous symmetry breaking. The roles played by this concept of generalized spontaneous symmetry breaking can be found, e.g., in \cite{Park:2013bma}\cite{Park:2017wiw}\cite{Park:2018xtt}.

%%%%%%%%%%%%%%%%%%%%%%%%%%%%%%
%%%%%%%%%%%%%%%%%%%%%%%%%%%%%%
\subsection{Flat spacetime analysis \la{gci}}
%%%%%%%%%%%%%%%%%%%%%%%%%%%%%%
%%%%%%%%%%%%%%%%%%%%%%%%%%%%%%

As stated in the beginning, the crux of our OPT is captured by considering a scalar system in a flat spacetime. We employ the $\overline{\mbox{MS}}$ subtraction scheme in the present section. We will come back to the deviation from the $\overline{\mbox{MS}}$ scheme in section 3.3.

%%%%%%%%%%%%%%%%%%%%%%%%%%%%%%
%\subsubsection*{OPT-induced renormalized mass}
%%%%%%%%%%%%%%%%%%%%%%%%%%%%%%

The starting point of the OPT-improved thermal resummation can be taken as the following renormalized action
\bea
S(\z) &=& -\int d^4x\; \fr12 \pa_\m \z \pa^\m \z  -\int d^4x\;\Big(\fr12 M^2 \z^2+ \fr{\l}{4!} \z^4\Big) -\int d^4x\;  \fr{3\n^4}{2\l}
\nn\\
% &=& -\int d^4x\; \fr12 \pa_\m \z \pa^\m \z  -\int d^4x\;\fr1{4!} \l \Big(\z^2 +\fr{6M^2}{\l}\Big)^2 -\int d^4x\;  \fr{3(\n^4-M^4)}{2\l}
%\nn\\ 
\la{esabrenm} 
\eea
with
\be
M^2(T)\equiv \n^2+\fr{\l}{24} T^2  \la{Massterm}
\ee
We take the expression $M^2(T)\equiv \n^2+\fr{\l}{24} T^2$ for the purpose of {\em computing the loop contributions}, namely, to take the thermal resummation into account. Later, we use $M^2\equiv \n^2$ for the {\em classical part} of the one-loop effective potential eq. \rf{1loopOPT}.\footnote{One may consider using the expression $M^2(T)\equiv \n^2+\fr{\l}{24} T^2$ even for the classical part of the one-loop potential in eq. \rf{1loopOPT}. This would amount to finite renormalization of the mass term. This possibility will be further examined in \cite{Park:2021vro}. \la{fnor}}
Shift the field 
\be
\z\ra \hat{\z}+\tilde{\z} \la{cs}
\ee
where $\hat{\z},\tilde{\z}$ denote the fluctuation field and background field, respectively. 
Since we are interested in the potential as opposed to the action, the background field $\tilde{\z}$ can be treated as a constant. Then the potential can be effectively computed by considering the field-dependent mass term
\be
M^2(T) \ra \Mt^2(T,\tilde{\z})= \n^2 +  \fr{\l}{24} T^2   +\fr{\l}{2}  \tilde{\z}^2
\la{Mmrel}
\ee
and integrating out the fluctuation field $\hat{\z}$. A remark is in order before proceeding to determination of the OPT-induced renormalized mass. We stated earlier that although we consider a CMB-order temperature, the high-temperature expansion can be utilized. The temperature being high or low is relative to the mass and we will show below that our OPT implies $\Mt\sim \hbar^{1/2} T$ ($\hbar$ will be kept implicit). One can now tell why the high-temperature expansion is justified: since the auxiliary mass $\Mt$ satisfies $\Mt/T\sim \hbar^{1/2}$, the intermediate analysis corresponds to that of high temperature: $\fr{\Mt}{T}<<1$.

The two-loop calculation of the effective potential was conducted long ago, e.g., in \cite{Parwani:1991gq}, \cite{Arnold:1992rz}, and \cite{Chiku:1998kd}. For our goal, it is necessary to keep tract of the field-independent terms as well. Also, the $M$- and $\Mt$- dependence is important. Let us focus on the one-loop potential; after carefully following these terms, one gets
\bea
V_{\mbox{opt}}(\tilde{\z})
&=&\fr{3\n^4}{2\l} -\fr{\pi^2T^4}{90}-\fr{\Mt^4}{32\pi^2}\ln\fr{\bar{\m}e^{\g_E}}{4\pi T} +\fr{1}{24} \Mt^2 T^2\nn\\
&&+\fr12\n^2\tilde{\z}^2-\fr{1}{12 \pi} \Mt^3T+\fr1{4!} \l \tilde{\z}^4+{\cal O}\Big(\fr{\Mt^6}{T^2}\Big) \la{1loopOPT}
\eea
where $\bar{\m}\equiv \m\Big(\fr{4\pi}{e^{\g_E}}\Big)^{1/2}$ is a scaling parameter of dimensional regularization (with the $\overline{\mbox{MS}}$ scheme). The field equation associated with \rf{1loopOPT}, $\fr{\pa }{\pa \tilde{\z}}V_{\mbox{opt}}=0$, yields
\bea
\fr{\l}{6}  \tilde{\z}^2+\n^2
-\fr{\l}{(4\pi)^2}\Mt^2\ln\fr{\bar{\m}e^{\g_E}}{4\pi T} +\fr{1}{24} \l T^2
- \fr{\l}{ 8\pi} \Mt T  =0
\eea
up to terms of two-loop order.
The solution is 
\bea
\tilde{\z}^2(M) 
%%%%%%%%%%%%%%%%%%%%%%
%%%%%%%%%%%%%%%%%%%%%%
&\simeq&-\frac{6 \n^2}{\l}-\frac{ T^2}{4}+\fr{3}{8\pi^2}\Big(-3\n^2+M^2\Big)\ln\Big(\fr{\bar{\mu}e^{\g_E}}{4\pi T}\Big)+\frac{3T \sqrt{   \left(M^2-3 \n^2\right)}}{4 \pi}+\cdots.\nn\\
\eea
With this, one gets the following onshell potential:
\bea
V_{\mbox{opt}}
&=&\fr{3\n^4}{2\l} -\fr{\pi^2T^4}{90}-\fr{\Mt^4}{32\pi^2}\ln\fr{\bar{\m}e^{\g_E}}{4\pi T} +\fr{1}{24} \Mt^2 T^2\nn\\
&&\hspace{-.9in}+ \fr12\n^2\tilde{\z}^2(M)-\fr{1}{12 \pi} \Mt^3T+\fr1{4!} \l \tilde{\z}^4(M)+{\cal O}\Big(\fr{\Mt^6}{T^2}\Big). \la{1loopOPT}
\eea
The following PMS condition for $\n$
\bea
\fr{\pa V_{\mbox{opt}}}{\pa \n}=0 \la{pVpn}
\eea
admits\footnote{Strictly speaking, one will have to keep $\n$ approaching zero to be compatible with renormalization group consideration. (This is a relatively minor point and does not, of course, qualitatively change the conclusions.)}
\bea
\n=0 \la{nuzero}
\eea
as a solution. This implies
\bea
{M}^2=\frac{1}{24}\l T^2 \la{icoeff}
\eea
Interestingly, this is the same as the result obtained in \cite{Blaizot:2014ffa} (see also \cite{Kneur:2015uha} and \cite{Kneur:2015moa}) by considering renormalization group and choosing appropriate renormalization conditions.\footnote{Two cautionary remarks: firstly, although we have been loosely speaking that the optimized renormalized mass is on the order of the temperature, it is really ${M}^2$ that is on the order of the temperature. (However, the other branches of the solutions of \rf{pVpn} have that property, and in this sense one may say that the solution is generically of the order of the temperature.) Secondly, in the earlier arXiv version (v1) of this work, there was an algebraic error. Although they were relatively minor, they rendered the coefficient in \rf{icoeff} different from $1/24$.}
The other branches of the solutions have undesirable features. For instance, in those branches the small $\tilde{M}$-expansion is not justified.
The condition \rf{nuzero} yields
\bea
\tilde{\z}^2&=&-\frac{ T^2}{4}+  i\frac{\sqrt{3\l}  }{8 \pi }T^2  +\cdots
\eea
Note the novelty that $\tilde{\z}^2$ takes a complex value. Once the potential \rf{1loopOPT} is evaluated with this value, one gets 
\bea 
&&V_{\mbox{opt}} =-\frac{1}{90}  \pi ^2 T^4+\cdots  \la{mainres}
\eea
which then allows one to avoid the fine-tuning problem. 

The complete two-loop offshell form of the potential will be presented in \cite{Park:2021vro}. There one again encounters the novelty: the potential develops an imaginary part, signaling instability of the vacuum toward zero temperature.\footnote{Strictly speaking, the potential itself remains real even at two-loop. However, this is because the source of the imaginary parts, $\Mt^{3}$, does not contribute even at two-loop. Due to the expected contribution of $\Mt^{3}$ term (and the terms with higher odd-integer powers of $\Mt$ that should appear in higher-loop computations), it is expected that the complexity of the potential will become manifest at three-loop and on.} More on this in the conclusion.

We briefly comment on the $T^4$-scaling of $V_{\mbox{opt}}$. This scaling is reminiscent of the scale parameter behavior of the radiation component in the $\L$CDM cosmology. Recall, however, that the scale parameter behaviors of the constituents of the $\L$CDM cosmology were obtained by treating $\L$ as a constant and at the same time assuming separated conservation of each component. There has been recent suggestions for considering a time-dependent $\L$ \cite{Riess:2016jrr}\cite{Oztas:2018jsu}. In particular, the possibility of "time-dependent or early dark energy" - both of which are consistent with the result \rf{mainres} and the observation made in the introduction on the CC of high temperature regime - was mentioned in \cite{Riess:2016jrr}. It should be worthwhile (and is being presently pursued in \cite{Park}) to carry out the analysis without imposing the separate conservation. This seems more natural, given the ``organic" nature of our Universe \cite{Park:2021vro}.

\subsection{Curved space analysis \la{csa}}

Whereas what we referred to as the second-layer perturbation in \cite{Park:2018vci} is necessary for the pure gravity sector computation, there exists, for the matter sector, a powerful shortcut based on the first-layer perturbation, the ``one-stroke" method. In the present work the first-layer perturbation will be used exclusively. From the results obtained, it becomes evident that the qualitative conclusion of the flat space analysis remains unchanged.

\subsubsection*{Graviton sector}

Let us recall the zero-temperature case first. In \cite{Park:2016zgt} and \cite{Park:2018vci}, we conducted the computation in a brute-force manner by employing the second-layer perturbation and viewing the classical CC as the graviton mass term. As shown, e.g., in \cite{Park:2018vci}, the result is a divergent CC term
\be
\sim \int \sqrt{-\gt}.
\ee
Strictly speaking, for a flat background, the one-loop results vanish in dimensional regularization in the absence of the classical CC treated as the graviton mass term. For consistency with the observed value of the CC, the classical (i.e., renormalized) CC will have to be set to $\sim T^4$ in the renormalization program described below, and thus will not qualitatively affect the proposed resolution of the CC problem.

The two-loop vanishes at zero temperature, due to the tracelessness of $P_{\m\n\r\s}$ and \rf{Pprop}. The results of $n$-loop with $n>2$ will presumably vanish in dimensional regularization due to the structures of the vacuum diagrams. Let us examine things in more detail. An arbitrary $n$-loop graph with $n\geq 2$ contains a product of vertices that can be written as
\bea
<h_{\a_1\a_2}h_{\a_3\a_4}h_{\a_5\a_6}\cdots h^{\b_1\b_2}h^{\b_3\b_4}h^{\b_5\b_6}\cdots>
\eea
where the upper and lower indices are fully contracted. Contractions of the fields lead to
\bea
\tilde{P}_{\m_1\n_1\r_1\s_1}\tilde{P}_{\m_2\n_2\r_2\s_2} \cdots 
\eea  
where again, all the indices are contracted one way or another with $\gt^{\m\n}$'s. Whenever a pair of $\tilde{P}$'s have a pair of the indices contracted, the explicit expression for $\tilde{P}$ given in \rf{fpt} can be used. One can also use \rf{Pprop} to reduce the total number of $\tilde{P}$'s until only one $\tilde{P}$ remains at the end. Since all of the indices must be contracted, the final expression must be $\sim \tilde{P}_{\m\n}{}^{\m\n}$. Thus the overall tensor structure gets simplified. As for the loop integrals, the two-loop diagrams, i.e., the sunset- and ``8" - types of diagrams vanish in dimensional regularization due to masslessness of a graviton. 
We expect the same to be true for higher loops. However, these diagrams will have non-vanishing contributions if either a CC term is present and treated as a graviton mass term or temperature enters. Further analysis of the finite temperature case will be presented in \cite{Park:2021vro}.

\subsubsection*{Matter-involving sector}

The matter-sector diagrams can be subdivided, depending on whether or not they involve a graviton loop. The matter-involving vertices do not, unlike the pure graviton vertices, come with $\fr{1}{\k^2}$. Since the graviton propagator comes with $\k^2$, the diagrams leading in $\k$ are those with a matter loop, which are our focus.

There exists a highly effective ``one-stroke" method of computing the matter-involving part of the effective action, in which the flat spacetime analysis can be entirely carried over. For this, note the explicit form of $\tilde{\D}(x_1-x_2)$ can be written as 
\bea
\tilde{\D}(x_1-x_2)=\int \fr{d^4k}{(2\pi)^4}\fr1{\sqrt{-\gt(x_1)}}\; \fr{e^{ik\cdot (x_1-x_2)}}{i (k_\m k_\n \gt^{\m\n}(x_1)+m^2)}. \la{fcp}
\eea
Defining ``flattened" momentum and coordinates as
\bea
K_{\underline{\a}}\equiv \et_{\underline{\a}}^\m\, k_\m\quad,\quad  X^{\underline{\b}} \equiv \et^{\underline{\b}}_\n \,x^\n  \quad,\quad  
\et_{\underline{\a}}^\m \et_{\underline{\b}}^\n\, \gt_{\m\n}=\eta_{\underline{\a}\underline{\b}}
\eea
where the underlined indices are flattened, one gets the flattened propagator:
\bea
\tilde{\D}(X_1-X_2)=\int \fr{ d^4K}{(2\pi)^4}\fr{e^{iK_{\underline{\g}} (X_1-X_2)^{\underline{\g}}}}{i (K_{\underline{\a}} K_{\underline{\b}} \h^{\underline{\a}\underline{\b}}+m^2)}.
\la{ffp}
\eea
When computing a diagram, one can pull out all of the background fields and contract the fluctuation fields. The propagators can then be transformed to the above. Afterward, the steps become parallel to those corresponding to the flat cases. The matter part of the effective action can thus be computed exactly in the same manner in which it is computed in the flat case.

%%%%%%%%%%%%%%%%%%%%%%%%%%%%%%
\subsection{Consistency of new subtraction scheme}
%%%%%%%%%%%%%%%%%%%%%%%%%%%%%%

\begin{figure}
	\hspace{-.7in}
	\centerline{
		\begin{minipage}[b]{5cm}
			\epsfxsize=9cm
			\epsfbox{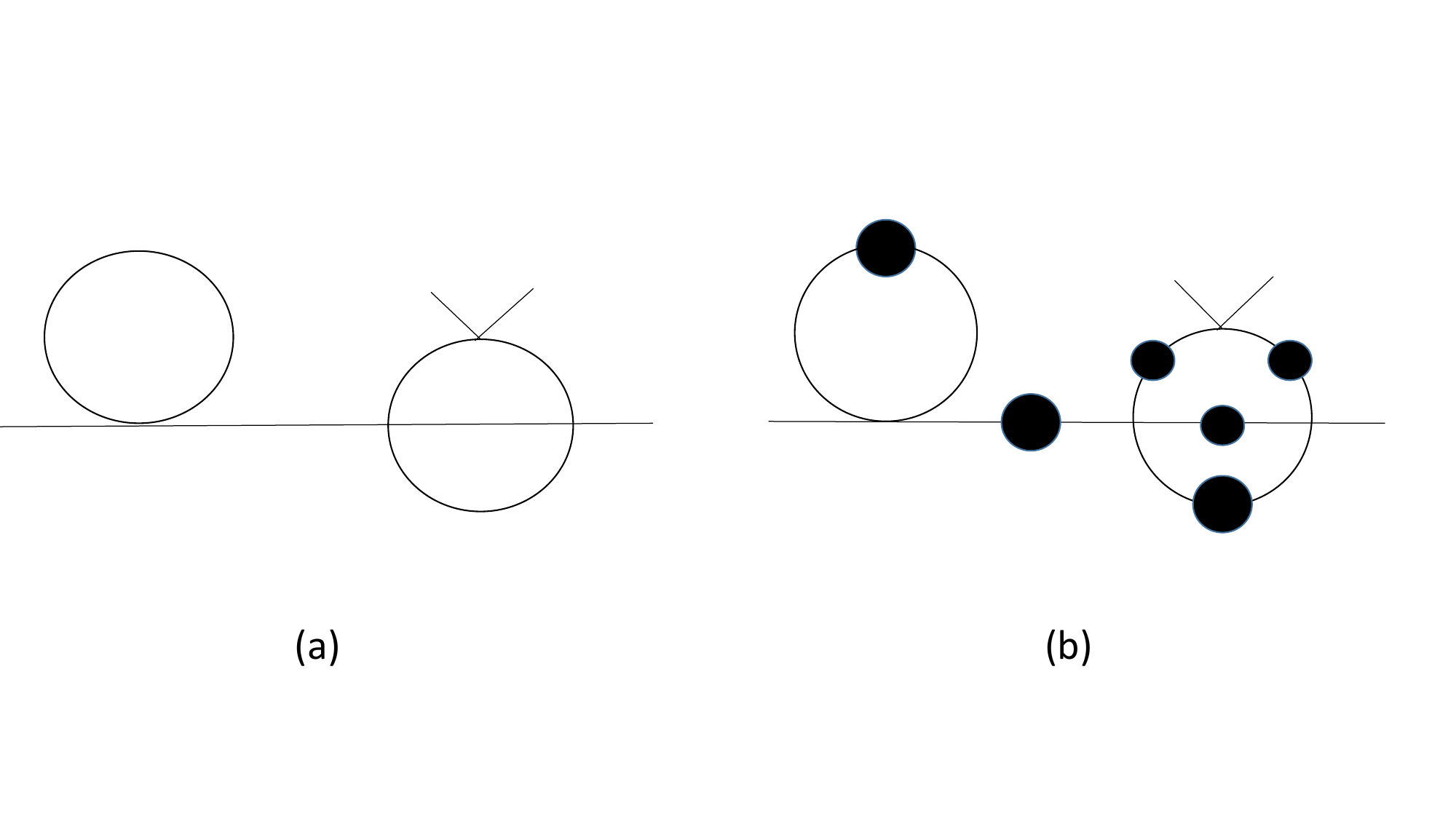}
		\end{minipage}
	}
	\vspace{-.3in}
	\caption{Resummation behind the new scheme: (a) a connected diagram in $\overline{\mbox{MS}}$ scheme (b) resummation required in the new scheme}
	\label{fig}
\end{figure}

With the renormalized mass around the CMB temperature, one should make sure that that framework preserves the success of the zero-temperature theory such as that seen in the zero-temperature SM. Again the success is guaranteed by bare perturbation theory. However, for practical purposes at the level of renormalized perturbation theory it will be illuminating to see how the steps of renormalized perturbation theory are modified in the new scheme. The new perturbation has been illustrated in section 2 by taking a simple scalar theory. In the case of the SM, each sector in the SM, i.e., the Higgs, gauge, and fermion, the renormalization process should be modified from that of the standard renormalization scheme, e.g., $\overline{\mbox{MS}}$. More specifically, in the modified scheme, the pole mass value must be realized by adjusting the finite parts of the divergent integrals that are chosen differently from those in the standard renormalization scheme: in analysis with $\overline{\mbox{MS}}$, the pole mass condition is part of the $\overline{\mbox{MS}}$ scheme, and the renormalized mass is not determined prior to the pole mass condition. It is the pole mass condition that determines the renormalized mass. In contrast, in the new scheme the renormalized mass is determined, as demonstrated in section 3.1, by the OPT. The finite parts are to be determined by the physical pole mass condition. What it implies is that the tree-level alone cannot achieve the accuracy of amplitudes obtained in $\overline{\mbox{MS}}$. In order to match the values of, e.g., the tree amplitudes computed in $\overline{\mbox{MS}}$, one must insert one-particle-irreducible diagrams to the internal lines. (As explained in the previous section, it will be useful to examine two-loop contributions to see how fast two-loop, compared to one-loop, decreases. We suggest this as a future task in the conclusion.) The insertions amount to a certain resummation. The situation is generally illustrated in Fig. 3. In light of this, it is also worth noting that the $T^2$-scaling of the renormalized mass was previously obtained in \cite{Blaizot:2014ffa} by choosing appropriate renormalization conditions.

%%%%%%%%%%%%%%%%%%%%%%%%%%%%%%
%%%%%%%%%%%%%%%%%%%%%%%%%%%%%%
\section{Conclusion}
%%%%%%%%%%%%%%%%%%%%%%%%%%%%%%
%%%%%%%%%%%%%%%%%%%%%%%%%%%%%%

Since some of the Standard Model particles, such as the Higgs, are massive, the matter contributions to the CC are naively expected to be larger than that of the graviton. The variant OPT reveals, however, that the temperature has an important input. This leads to the necessity of employing a new renormalization scheme. In the new scheme, one needs to go higher loop orders and perform a certain resummation to achieve the same level of proximity to the values of the physical observables as in the zero-temperature standard schemes. Given what it brings, this seems to be a relatively small price to pay.

The finding in the present work that a non-traditional renormalization scheme is needed to tackle the cosmological constant problem is in line with the observation, e.g., in \cite{Shapiro:2000dz}. The overarching umbrella for this and \cite{Shapiro:1999zt,Sola:2007sv,Sola:2016jky,Peracaula:2018dkg,Moreno-Pulido:2020anb} is the so-called running vacuum with time-varying and dynamical energy, a view that we find very natural. In these works, the cosmological constant problem was coherently tackled from different angles and various cosmological implications of such a vacuum were discussed. (Even the relevance of temperature to the cosmological constant problem was discussed to some extent.)
Several different regularization methods and their pertinent renormalization schemes were analyzed. We believe that the present work lends support, by providing independent motivation to consider a non-standard renormalization scheme, to those approaches. It is also interesting to note that the present finite temperature contribution to the stress-energy tensor seems to bear a certain resemblance to the `stress-energy tensor of cosmic interaction' obtained in \cite{Ryskin:2014pva}.

Before commenting on potential implications of the new scheme for cosmology, let us discuss some checks that are desirable for solidifying the new renormalization scheme in the context of renormalized perturbation theory. A potentially far-reaching peculiarity of the present scheme worth a closer look is that the one-loop part (or the total sum of the quantum corrections, to be exact; see footnote \ref{fn9}) should be bigger than the “classical part". (See the last paragraph for the reason for the quotation marks.) This is because the renormalized mass is taken to be far smaller than the physical value. This is the point illustrated in Fig. 3: since the pole mass in the new scheme will match poorly with the physical value at the classical level, one should go to at least one-loop and in general higher loops. It should be useful to look into things in two directions: Firstly, it will be illuminating to explicitly check how the new scheme will play out in an explicit SM example. Secondly, it will be useful to examine in the new scheme the rate at which the two-loop decreases compared to one-loop  and how sensitively the rate depends on the finite constants $c$'s. In the case of a scalar theory, one should be able to check this with reasonable amount of effort. These directions should be highly useful exercises in renormalized perturbation theory with the present novel scheme.

 Several ramifications of our results and further directions are in order. The results obtained in this work suggest that, earlier in the thermal history of the Universe, the value of the CC should have been larger. In other words the CC becomes time-dependent through the temperature, and the present small value must be due to the age of the Universe\cite{Weinberg}. More quantitatively, the following is the basis for this anticipation. In \cite{Park:2016zgt} it was shown that there exists a time-dependent solution that approaches the minimal value of the potential. By using the one-parameter family of the potentials labeled by the temperature, one can repeat the analysis with the present finite temperature potential. One can then analyze the resulting quantum-level action, and it should be possible, at the time-dependent solution level, to establish a CC that decreases to a small value. One may introduce a renormalized CC, $\L_{ren}$. To be consistent with the fact that the observed value of the CC is small, one will have to take $\L_{ren}$ to be small. It will be of interest to pursue this line of study. 
 
Another ramification, not unrelated to the first, is that the large value of the CC becomes natural when the temperature is on the order of the EW scale. (This may be something profound, and have additional implications for the hierarchy problem in the SM.) Since the Universe was at higher temperatures in the previous eras, it will be a meaningful endeavor to explore whether one could come up with a streamlined description covering the entire temperature range, say, from the near-Planck era to the present. With the recent progress in the OPT literature, the present results indicate toward an affirmative answer. Also, it will be interesting to study, e.g., the implications of the temperature effects for Hubble tension and early dark energy proposals \cite{Riess:2016jrr,Poulin:2018cxd}. In the early dark energy proposals it has been noted that the presence of extra dark energy before recombination has an effect of boosting the value of the present Hubble constant, thus lessening the tension. The finite temperature effects must be important in the post-inflationary reheating period. Soon after the period, equilibrium thermodynamic should be applicable. The present result implies that there must have been large time-dependent vacuum energy from very early on. We will report on the findings of further study in relation to the early dark energy proposal elsewhere \cite{Park}.

Lastly, as we will report in \cite{Park:2021vro}, the potential is expected to develop a imaginary part. We interpret this as an indication of the vacuum decay from a finite temperature to zero temperature. Just as an analogous decay induced by a bounce worm hole solution plays an important role in black hole information \cite{Park:2017wiw}\cite{Park:2019lbj}, the present vacuum is likely to have interesting cosmological implications that deserve further study.

We end with a highly speculative idea along the line of the discussion at the end of section 2 where we explored a possible connection between the peculiarity and the SM hierarchy problem. Conventionally, quantum effects are regarded as small. However, it is also well known that there exist phenomena, such as superfluidity, that are at odds with the smallness in that they manifest macroscopically. (See also \cite{Park:2017dib}, \cite{Nurmagambetov:2018het}, \cite{Nurmagambetov:2020ann} for the proposal of a quantum origin of astronomical jets.) One of the implications of the present work is the possibility that the dichotomy between classical and quantum physics is not as meaningful as used to be believed. In other words, since the splitting between the renormalized Lagrangian and quantum corrections are arbitrary, one may adopt the new scheme in all sectors. Normally, there is no need for taking such an extreme measure. However, it is noteworthy that only this kind of approach allows one to avoid the CC problem and at the same time preserve the success of the SM (up to the checks suggested above). In this sense it may be said that mass has a quantum origin.

\newpage
%%%%%%%%%%%%%%%%%%%%%%%%%%%%%%%%%%%%%%%%%%%%%%%%%%%%%%%%%%%%%%%%

\end{document}